\documentclass[12pt,a4paper,leqno]{amsart}
\usepackage{latexsym, amsfonts, amsthm, amsmath, amssymb, url, color,
  graphicx, epsfig, float} 
\usepackage[mathscr]{euscript}

\usepackage[bookmarks,bookmarksopen,colorlinks]{hyperref}
 \makeatletter
 \newif\ifmsbmloaded@

\input{scrload}

\title{Hidden Polynomial(s) Cryptosystems}
\author{\href{mailto:toli@posso.dm.unipi.it}{\texttt{Ilia Toli}}}
\address{{Dipartimento di Matematica
  {\it Leonida Tonelli}\\ via F. Buonarroti 2\\ 56127 Pisa\\
  Italy.\\\href{mailto:toli@posso.dm.unipi.it}
  {\texttt{\tt toli@posso.dm.unipi.it}}}}
\date{}
\begin{document}
\keywords{Public key, hidden monomial, HFE, differential algebra, TTM,
probabilistic encryption, signcryption, private key.} 
\subjclass{Primary: 11T71; Secondary: 12H05}
\begin{abstract}We propose public-key cryptosystems with public key a
  system of polynomial equations, algebraic or differential, and
  private key a single polynomial or a small-size ideal. We set up
  probabilistic encryption, signature, and signcryption
  protocols.\end{abstract}         
\maketitle
\section{Introduction}
This paper focuses on Hidden Monomial Cryptosystems, a class of
public key (PK) cryptosystems first proposed by Imai and
Matsumoto \cite{imai1}.  In this class, the
PK is a system of polynomial nonlinear equations. The private key
is the set of parameters that the user chooses to construct the equations.
Before we discuss our variations, we review
briefly a simplified version of the original cryptosystem, better
described in \cite{koblitz}. 
The parties throughout this paper are:
\begin{itemize}
\item Alice who wants to receive secure messages;
\item Bob who wants to send her secure messages;
\item Eve, the eavesdropper. \end{itemize}

Alice takes two finite fields $\mathbb{F}_q<\mathbb{K}$, $q$ a power of
$2$, and $\beta_1, \beta_2, \ldots , \beta_n$ a basis of
$\mathbb{K}$ as an $\mathbb{F}_q$-vector space. Next she takes $0<h<q^n$
such that $h=q^{\theta}+1$, and $gcd(h,q^n-1)=1$. Then she takes two
generic vectors ${\bf
  u}=(u_1,\ldots,u_n)$ and ${\bf v}=(v_1,\ldots,v_n)$ upon $\mathbb{F}_q$, and
sets\footnote{In this paper we reserve {\bf boldface}
   to the 
  elements of $\mathbb{K}$ thought as vectors upon $\mathbb{F}_q$ in
  the fixed private basis. They are considered vectors or field
  elements, as convenient, without further 
  notice. This shift in practice takes a Chinese Remainder Theorem.
 {\it  Cryptosystem} and {\it scheme} are synonyms.}:  
\begin{equation}{\bf   
  v=u}^{q^{\theta}} {\bf u}.\label{vuu}\end{equation} 

 The condition $gcd(h,q^n-1)=1$ is equivalent to requiring that the map ${\bf
  u}\longmapsto~{\bf u}^h$ on $\mathbb{K}$ is  ${\it
  1\!\!\leftrightarrow\!\!1}$; 
  its inverse  
  is the map ${\bf u}\longmapsto {\bf u}^{h'},$ where $h'$ is the
  inverse multiplicative of $h$ modulo $q^n-1$.

In addition, Alice chooses two secret affine transformations, i.e.,
two invertible matrices $A=\{A_{ij}\}$ and $B=\{B_{ij}\}$ with entries
in $\mathbb{F}_q$, and two constant vectors ${\bf c}=(c_1,\ldots,c_n)$
and ${\bf d}=(d_1,\ldots,d_n)$, and sets:
\begin{equation}{\bf u}=A{\bf x+c}\qquad \mbox{and} \qquad {\bf v}=B{\bf
    y+d}.\label{aff}\end{equation} 

 Recall that the operation of raising to the
$q^k$-th power in $\mathbb{K}$ is an $\mathbb{F}_q$-linear
transformation.
Let $P^{(k)}=\{p_{ij}^{(k)}\}$ be the matrix of this
linear transformation in the basis $\beta_1, \beta_2, \ldots ,\beta_n$, i.e.:
\begin{equation} 
\beta_i^{q^k}=\sum_{j=1}^n p_{ij}^{(k)}\beta_j, \qquad
p_{ij}^{(k)}\in\mathbb{F}_q , \label{id1}
\end{equation}
for $1\leq i,k\leq n$. Alice also writes all products of basis elements
in terms of the basis, i.e.:
\begin{equation} 
\beta_i\beta_j=\sum_{\ell=1}^n m_{ij\ell}\beta_{\ell}, \qquad m_{ij\ell}\in
\mathbb{F}_q, 
\label{id2}\end{equation}
for each $1\leq i,j\leq n$. 
Now she expands the equation (\ref{vuu}). So she obtains a system of
equations, explicit in the $v$, and quadratic in the $u$. She uses now
her affine relations (\ref{aff}) to replace the $u,v$ by the
$x,y$. So she obtains $n$ equations, linear in the $y$, and of degree
$2$ in the $x$. Using linear algebra, she can get $n$ explicit
equations, one for each $y$ as polynomials of degree $2$ in the $x$.

Alice makes these equations public. Bob to send her a message $(x_1,
x_2, \ldots ,x_n)$, 
substitutes it into the public equations. So he obtains a linear system of
equations in the $y$. He solves it, and sends  ${\bf y}=(y_1,
y_2,\ldots,y_n)$ to Alice. 

To eavesdrop, Eve has to substitute
$(y_1,y_2, \ldots ,y_n)$ into the public equations, and solve the
nonlinear system of equations for the unknowns $x$.

When Alice receives {\bf y}, she decrypts:
\begin{eqnarray*}&y_1, y_2,\ldots,y_n&\\
&\Downarrow&\\
&{\bf v}=B{\bf y+d}&\\
&\Downarrow&\\
&{\bf v}=\sum v_i\beta_i &\\
&\Downarrow&\\
&{\bf u=v}^{h'}&\\
&\Downarrow&\\
&{\bf x}=A^{-1}({\bf u-c}).&
\end{eqnarray*}

In Eurocrypt $'88$ \cite{imai2}, Imai and Matsumoto proposed a digital
signature algorithm for their cryptosystem. 

At Crypto $'95$, Jacques Patarin \cite{Patarin95} showed how to break this
cryptosystem. He noticed that if one takes the equation  ${\bf
  v=u}^{q^{\theta}  +1}$, raises both sides on the $(q^{\theta}-~1)$-th
power, and multiplies both sides by ${\bf uv}$, he gets the equation ${\bf
  u v}^{q^{\theta}}={\bf u}^{q^{2\theta}} {\bf v}$ that
leads to equations in the $x$, $y$, linear in both sets of
variables. Essentially the equations do not suffice to identify uniquely
the message, but now even an exhaustive search will be
feasible. The system was definitively insecure and breakable, but its
ideas inspired a whole class of PK cryptosystems and digital
signatures based on structural identities for finite field operations
\cite{HFE, moh, koblitz, Patarin96, patarin96hidden, gou-pat1}.   

The security of this class rests on the difficulty of the
problem of solving systems of nonlinear polynomial equations. This problem is
hard iff the equations are randomly chosen. If they really were
random, the problem is hard to Alice, too. So, all we try to do is to get
systems of equations that are not random, but appear to be the most possible. 

Our paper is organized as follows. In the next section we develop an our
  own, new cryptosystem. Alice builds her PK by manipulations
  as above, starting from a certain bivariate polynomial. 

All of
  Alice's manipulations are meant to hide from Eve this polynomial. It
  is the most important part of the private key. Its knowledge reduces
  decryption to the relatively easy problem of solving a single
  univariate polynomial of a moderate degree.

Encryption is probabilistic, in the sense that to
a given cleartext correspond zero, one, or more
ciphertexts. Decryption is deterministic, in the sense that if
encryption succeeds, decryption does succeed, too.

Almost any bivariate nonlinear
polynomial can give raise to a PK. This
plentitude of choices is an important security parameter.

In the third section we discuss some security issues.
In the fourth one we provide our cryptosystem with a digital
signature algorithm. 

In the fifth we provide a signcryption protocol. {\it Signcryption}
stands for {\it joint encryption and signature}.

In the sixth one we discuss some more variations. Essentially, we
replace the single bivariate polynomial by an ideal of a small size.

In the seventh section we mention what Shannon \cite{stinson} calls
  {\it unconditionally secure cryptosystems.} Nowadays they are
  considered an exclusive domain of the private key cryptography. This
  is due mostly to the unhappy state of art of the PK one. 

In the eighth one we extend our constructions to differential fields
of positive characteristic. We hope they are the suitable environment
for unconditionally secure PK (USPK) cryptosystems.
\section{A New Cryptosystem}
\subsection{Key Generation}\label{key}
Alice chooses two finite fields
 $\mathbb{F}_q <\mathbb{K}$,  
 and a basis $\beta_1, \beta_2,\ldots, \beta_n $  of
 $\mathbb{K}$ as an  $\mathbb{F}_q$-vector space. In practice,
 $q=2$. However, it can be any $p^r$, for any $p$ prime, and any
 $r\in\mathbb{N}$. 

Next Alice
takes a generic (for now) bivariate polynomial:
\begin{equation}f(X,Y)=\sum_{ij}{{\bf a}_{ij}X^iY^j\label{poly1}}\end{equation}
in $\mathbb{K}[X,Y]$, such that she is able to find {\bf all} its roots in
$\mathbb{K}$ with respect to $X$; $\forall$ $Y \in \mathbb{K}$, if any. 
For the range of $i$ employed, this is nowadays considered a relatively
easy problem. Further, $f(X,Y)$ is subject to other few constraints, that
 we make clear at the opportune moment.

In transforming cleartext into ciphertext message, Alice will work
with two intermediate vectors, ${\bf u}=(u_1,\ldots,u_n)$ and ${\bf
  v}=(v_1,\ldots,v_n)$; ${\bf u, v \in \mathbb{K}}$. 
She sets: 
\begin{equation}
\sum_{ij}{{\bf a}_{ij}{\bf u}^i{\bf
      v}^j}=0. \label{poly} \end{equation}  
 
For ${\bf a}_{ij} \neq 0$, she sets somehow: 
 \begin{equation} 
i=\sum_{k=1}^{n_{i}} q^{\theta_{ik}}\qquad\mbox{and}\qquad
j=\sum_{k=1}^{n_{j}} q^{\theta_{jk}}, 
\label{equal}\end{equation}
where $\theta_{ik}, \ \theta_{jk}, \  n_{i}, \ n_j, \ i, \ j \
\in\mathbb{N}_*=\{0,1,2,\dots\}$.  

Here {\it somehow} means that (\ref{equal}) may or may not be the $q$-ary
representation of $i$, $j$. Taking this freedom, we
increase our range of choices, whence the random-looking of the PK. In
any fashion, what we are dealing with, are nothing but identities.

Next Alice substitutes the (\ref{equal}) to the exponents in
(\ref{poly}), obtaining:
\begin{equation}
\sum_{ij}({{\bf a}_{ij} exp({\bf u},{\sum_{k=1}^{n_i}
  q^{\theta_{ik}}}) exp({\bf
  v},{\sum_{k=1}^{n_0} 
  q^{\theta_{jk}}})})=0;
\end{equation} 
that is:
\begin{equation}
\sum_{ij}({{\bf a}_{ij} \prod_{k=1}^{n_i}{\bf u}^{
  q^{\theta_{ik}}}}\prod_{k=1}^{n_j}{\bf v}^{
  q^{\theta_{jk}}}) =0.
\label{prod}\end{equation}

{\bf Recall that the operation of raising to the
$q^k$-th power in $\mathbb{K}$ is an $\mathbb{F}_q$-linear
transformation.} 
Let $P^{(k)}=\{p_{\ell m}^{(k)}\}$ be the matrix of this
linear transformation in the basis $\beta_1, \beta_2, \ldots ,\beta_n$, i.e.:
\begin{equation} 
\beta_{i}^{q^k}=\sum_{j=1}^n p_{ij}^{(k)}\beta_j, \qquad
p_{ij}^{(k)}\in\mathbb{F}_q ; \label{id3}
\end{equation}
for $1\leq i,\,j\leq n$. Alice also writes all products of basis elements
in terms of the basis, i.e.:
\begin{equation} 
\beta_{i}\beta_j=\sum_{k=1}^n m_{ijk}\beta_{k}, \qquad
m_{ijk}\in\mathbb{F}_q; 
\label{id4}\end{equation}
for $1\leq i,\,j\leq n$. 

Now she  substitutes ${\bf u}=(u_1, u_2,\ldots,u_n)$, ${\bf a}_{ij}=(a_{ij1},
a_{ij2},\ldots,a_{ijn})$,
${\bf v}=(v_1,v_2,\ldots,v_n)$, and the
identities (\ref{id3}), (\ref{id4}) to (\ref{prod}), and
expands. So she 
obtains a system of $n$ equations of degree $t$ in
the $u$, $v$, where:
\begin{equation}t\ =\ max \  \{n_{i}+n_j\ \ :\ \
   {\bf a}_{ij}\neq 0\}.\label{set}\end{equation} 

Every term under the $\Sigma$ in (\ref{equal}) contributes by one to the
size of $t$.

Here we pause to give some constraints on the range of $i$, $j$ in
(\ref{poly}). The 
aim of this section is to generate a set of polynomials; linear in a
set of variables, and nonlinear in another one. For that purpose, we
relate (\ref{poly}) and (\ref{equal}): ${\bf a}_{ij}\neq 0
\Rightarrow$ $\{n_i>1$, $n_j=1\}$.

On the other side, the size of PK is
$\mathcal{O}(n^{t+1})$. So, it grows polynomially with $n$, and
exponentially with $t$. Therefore, we are interested to keep $t$
rather modest, e.g., $t=2,\ 3$, or so. So, we
have to choose $i$, $j$ in (\ref{poly1}), (\ref{equal})  in order to
keep $t$ under a forefixed bound.

Next she takes $A=\{A_{ij}\}, B=\{B_{ij}\}\in GL(\mathbb{F}_q)$,
${\bf c}, {\bf d}\in\mathbb{K}$, and sets: 
\begin{equation}
{\bf u}=A{\bf x+c} \qquad\mbox{and}\qquad {\bf v}=B{\bf y+d}, \label{matrix}
\end{equation}
where ${\bf x}=(x_1,x_2,\ldots,x_n)$, ${\bf y}=(y_1,y_2,\ldots,y_n)$ are
vectors of variables.

Now she substitutes  (\ref{matrix}) to the equations in the $u$,
$v$ above, and expands. So she  
obtains a system of $n$ equations of degree $t$ in the $x$, $y$;
linear in the $y$, and nonlinear in the $x$.

After the  (\ref{matrix}) each monomial $X_iY_j$
expands into polynomials with terms of each degree, 
from $n_i+n_j$ to zero. 
So, they  shuffle better the terms coming from different
monomials of (\ref{prod}). On the other hand, they render the PK very
dense, so increase drastically its size. 

At this point, we are ready to define the cryptosystem. 
\subsection{The Protocol}With the notations adopted above, we
define\label{protocol} the {\bf HPE 
  Cryptosystem} (Hidden Polynomial Equations) as the PK
  \mbox{cryptosystem} such that:
\begin{itemize}
\item {\bf The public key is:}
\begin{itemize}\item The set of the polynomial
    equations in the $x$, $y$ as above;
\item The field $\mathbb{F}_q$;
\item The alphabet: a set of elements of $\mathbb{F}_q$, or strings of them.
\end{itemize}
\item {\bf The private key is:} \begin{itemize}
\item The polynomial (\ref{poly1});
\item $A$, $B$, ${\bf c}$, ${\bf d}$ as in (\ref{matrix}); 
\item The identities (\ref{poly}) to (\ref{id4});
\item The field $\mathbb{K}$.
\end{itemize}
\item {\bf Encryption:}
Bob substitutes the cleartext
${\bf x}=(x_1,x_2,\ldots,x_n)$ in the 
public equations, solves with respect to the $y$, and sends ${\bf
  y}=(y_1,y_2,\ldots,y_n)$ to Alice. We assume that
solutions exist, and postpone the case when there are not.  
\item  {\bf Decryption:}
Alice substitutes
  ${\bf v}=B{\bf y+d}\in\mathbb{K}>\mathbb{F}_q$ in 
(\ref{poly}), and finds {\bf all} solutions within $\mathbb{K}$.  
There is at least one. Indeed, if ${\bf x}$ is Bob's cleartext, ${\bf
  u}$ as in (\ref{matrix}) is one. 
For each solution ${\bf u}$, she solves:
  \begin{equation}{\bf x}=A^{-1}({\bf u-c}),
  \label{expl}\end{equation}and represents all solutions in the basis
  $\beta_1, \beta_2,\ldots, \beta_n $. It takes a Chinese Remainder
  Theorem. With probability $\approx 1$, all 
results but one, Bob's $(x_1,x_2,\ldots,x_n)$, are gibberish, or even stretch
out of the alphabet. We come back later at this point, too. 
\end{itemize} 
\subsubsection{}The main suspended question is
  that of the existence of 
  solutions. Well, Bob succeeds to encrypt a certain message {\bf x}
  iff Alice's equation (\ref{poly}) has solutions for {\bf u} as in
  (\ref{matrix}) for that {\bf x}. Alice's polynomial (\ref{poly}) in
  {\bf v} for a given {\bf u} is a random
  one. It is a well-known fact from algebra that the
probability that a random polynomial with coefficients upon a finite
field has a root in it is
$1-\frac{1}{e}\approx 63.2\%$ \cite{koblitz, marcus}. 
\label{remedy}

Here the remedy is probabilistic. Alice renders the alphabet public
with letters being sets of elements of $\mathbb{F}_q$, or sets of strings in
it. Bob writes down a plaintext,
and starts encryption. If he fails, he substitutes a letter or a
string of
the cleartext with another one of the same set, and retries.
After $s$ trials, the probability 
he does not succeed is $\frac{1}{e^s}$; practically good enough.
\subsubsection{}The other problem is that Alice may have to
distinguish the right solution among a great number of them. Here is
 a first remedy. Her number of solution is bounded above by the
degree in $X$ of $f$. So, it is beter to keep it moderate. 
Later we give other remedies, too.

\subsection{Observations} Solving univariate polynomial equations is used by
  Pa\-ta\-rin, too \cite{patarin96hidden, Wolf:02:Thesis}. He takes a
  univariate polynomial:
  $$f(x)=\sum_{i,j}\beta_{ij}x^{q^{\theta_{ij}}+q^{\varphi_{ij}}}+
  \sum_i\alpha_ix^{q^{\xi_i}}+\mu_0,$$
and with manipulations like ours, both the same as Imai-Matsumoto
  \cite{imai1}, he gets his PK; a set of
  quadratic equations. He uses two
  affine transformations to shuffle the equations. We claim that the
  first one adds nothing to the security.

The bigger the degree of $f$ is, the more the PK resembles a
  randomly chosen set of quadratic equations. So, it is a security
  parameter.  On the other side, it slows down decryption, principally
  by adding a 
  lot of undesired solutions. To face that second problem, to the
  PK are added other, randomly chosen, equations. This is its
  {\it Achilles' heel}. It
  makes the PK overdefined, therefore subject to certain
  facilities to solve \cite{ckps}. So, it weakens the trapdoor
  problem.

We do not add equations to discard
undesired solutions. Indeed, we take the degree in $X$ rather modest,
so we do not have so many undesired solutions.
Thus, we are not subject to attacks exploiting overdefined
equations. If in certain variations 
we ever do, we need to add less equations, however.

What is most important, we have now a practically infinite range of choices of
$f$. This is not Patarin's case. There the choices are bounded below
because of being easy to attack cases, and above because of being impractical
to the legitimate users.

The only few constraints we put on monomials of $f$ aim to:
\begin{itemize}
\item keep PK equations linear in the $y$; 
\item have less undesired solutions in decryption process;
\item keep the  size of PK moderate;
\item keep {\bf all} PK equations nonlinear in the $x$.
\end{itemize}\label{bivar}

The constraint that {\bf all} PK equations {\bf must} be
nonlinear in the $x$ is the only non-negotiable one. Indeed, if Alice
violates it, the trapdoor problem becomes fatally easy to Gr\"obner
techniques.

 We can take the degree in {\bf y} arbitrarily huge. It 
  gives no trouble to us. We only require the monomials of $f$ to be
  of the form ${\bf x}^i {\bf y}^{q^j}$ for $i,j\in\mathbb{N}_*$, so the
  public equations come linear with respect to the $y$. 

Assume now that PK is nonlinear in the $y$. Once Bob
substitutes the $x$ in the public equations, he is required to {\bf
  find any solution} of the system that he obtains. This can be done
within polynomial time with respect to Bezout number of the
system. Later we give settings to keep PK nonlinear of low total
degree in the $y$. 

Each of such solutions (if any) is encryption to the same cleartext. So
we have set up a probabilistic encryption protocol. To a single cleartext
may correspond zero, one, or more ciphertexts.

\section{Security Issues}
The main data to Eve are the system of public equations and the 
order of extension. By brute force, she has to take
$(y_1,y_2,\ldots,y_n)$, to substitute it in the PK equations, to
solve within the base field, and to take the sensate
solution. Almost surely, 
there is only one sensate solution among those that she finds.
 She has to find it among $t^n$ of them.
 However, the  main difficulty
 to her is just 
 solving the system. Supposedly, it will pass through the complete
 calculus 
 of a Gr\"obner basis. It is a well-known hard problem. 

So, the complexity of the trapdoor problem is $\mathcal{O}(t^n)$.
On the other hand, the size of the PK is
$\mathcal{O}(n^{t+1})$. This fully suggests the 
values of the parameters. It is better to take  
$n$ huge. This diminishes the probability that Alice confuses decryption,
however close to zero, and, what is most important, it renders Eve's
task harder. Alice and Bob will have to solve sets of bigger systems of
 linear equations, and face Chinese Remainder Theorem for bigger $n$.

If we take $t$ very small, we restrict somehow choices of $f$. If very
big, it renders 
the size of PK impractical. Actually, $n\geq 100$  and $t=2,3,4$ are
quite good
sample values. If we only take the monomials of $f$ to be univariate,
PK size is roughly the same as $HFE$, and we have infinite choices
still. In any case, later in section \ref{ideal} we present better
settings that all in one: moderate the size of the PK, increase its
randomness, and contain better the number of undesired solutions.

There exist well-known facilities \cite{ckps} to solve overdefined systems of
equations. Unlike most of the rest, our PK is irrendundant, so
it is not subject to such facilities.

Now, by exhaustive search we mean that Eve substitutes the $y$ in the
public equations, and tries to solve it by substituting values to the $x$.
If we have $d$ letters each of them being represented by a single
element of $\mathbb{F}_q$, the complexity of an exhaustive search is
$\mathcal{O}(d^n)$. It is easy for Alice to render exhaustive search
more cumbersome than 
Gr\"obner attack. The last one seems to be the only choice to Eve.

{\it Affine multiple attack}  \cite{patarin96hidden} seems
of no use in these settings.

Obviously, infinitely many bivariate polynomials give raise to the same public
key. Indeed, fixed the ground field, the degree of extension $n$, and
the degree of PK equations, we have a finite number of public
keys. On the other hand, there are infinitely many bivariate polynomials that
can be used like private keys. 

On how does it happen, nothing is known. If ever found, any such 
regularity will only weaken the trapdoor problem.
\section{A Digital Signature Algorithm}\label{sign}
For Bob to be able to sign messages, he builds a
cryptosystem as above with $[\mathbb{K}_B:\mathbb{F}_{q_B}]=n_B$.
Assume now that we are publicly given a set of hash functions that send
cleartexts to $n_B$-tuples of $\mathbb{F}_{q_B}$. 

Bob to sign a message $M$: 
\begin{itemize}
\item calculates $H(M)=(y_1,y_2,\ldots,y_{n_B})={\bf y}_B$, then ${\bf
    v}_B=B_B{\bf y}_B+{\bf d}_B$; 
\item finds one solution (if any; otherwise, see section
  \ref{remedy}) ${\bf u}_B$ of $f_B({\bf u}_B,{\bf v}_B)=0$ in $\mathbb{K}_B$.
\item calculates ${\bf x}={A_B}^{-1}({\bf u}_B-{\bf c}_B)$;
\item appends ${\bf x}=(x_1,x_2,\ldots,x_{n_B})$ to $M$, encrypts,
 and sends it 
 to Alice.  $(x_1,x_2,\ldots,x_{n_B})$ is a signature to $M$.\end{itemize}  
 
To authenticate, Alice first decrypts, then she calculates
$H(M)=(y_1,y_2,\ldots,y_{n_B})$.
If $(x_1,x_2,\ldots,x_{n_B})$, $(y_1,y_2,\ldots,y_{n_B})$ is a
  solution of Bob's PK, she accepts the message; otherwise she knows
  that Eve has been causing trouble.

If Eve tries to impersonate Bob and send to Alice her own message with hash
value ${\bf y}=(y_1,y_2,\ldots,y_{n_B})$, then to find a signature
$(x_1,x_2,\ldots,x_{n_B})$, she may try to find one solution of Bob's system
of equations for {\bf y}.
We trust on the hardness of this problem for the security of
authentication.

Actually, the hash functions play no role in this class of
signatures. They may as well output parts of the cleartext itself.

\section{A Signcryption Protocol}\label{pep}

Here is the shortest possible description. Let $F_A$ and $F_B$ be
Alice's and Bob's PK functions respectively. To send
a message {\bf x} to Alice, Bob sends her a random element of
$F_A(F^{-1}_B({\bf x}))$,
that she can decrypt by calculating 
$F_B(F^{-1}_A(F_A(F^{-1}_B({\bf x}))))$. 
So if $F_A(F^{-1}_B({\bf
  x}))\neq \emptyset$. Otherwise, the approach is probabilistic, as in
the previous section.

Here is the extended description. Each
letter (or some of them, only) is represented by a set of few
(two, e.g.) elements of the field, or 
strings of them. For ease of explanation, assume that
$\mathbb{F}_{q_A}=\mathbb{F}_{q_B}$ and $n_A=n_B$.

Bob writes down the cleartext $X$, calculates ${\bf v}_B=B_BX+{\bf
    d}_B$, and finds one solution (if any, otherwise see section
    \ref{remedy}) ${\bf u}_B$ of his private polynomial $f_B(X,Y)$. 
Next he calculates ${\bf x}_B={A_B}^{-1}({\bf u}_B-{\bf c}_B)$, that
    he encrypts as above by means of Alice's PK, and sends her the
    result.

Alice now first decrypts as in section \ref{protocol}. Next, she
substitutes the {\bf x}-es she finds into Bob's PK variables $x$, and
solves. There 
is at least one solution, and at most few of them. One of them is
Bob's message. 

What is the trapdoor problem now?
Well, on authentication matter, nothing new. Eve has the same chances
to forge here that she had before. Recall that this class of signatures
is already considered best with respect to the other ones.

On security, instead, there is a very good improvement. By brute
force, Eve has to take the 
ciphertext, substitute on Alice's PK, find all solutions,
substitute them all on Bob's PK, and take the sensate ones. 

Let us assume that the letters are strings of a fixed length. For an
exhaustive search Eve now has 
to run throughout all the $n$-tuples of all elements of Alice's ground field;
not just throughout $n$-tuples made of letters.
She sets up such $n$-tuples, checks whether
they are solutions of Alice's PK for Bob's ciphertext
{\bf y} substituted to the variables $y$. If yes, she substitutes to
Bob's PK, and takes the sensate ones.

So, Alice now has a full freedom on building alphabet. In
decryption she discards a priori the solutions that contain
non-letters. Now practically the good solution is unique.

Apart all, we save the space and calculi of the signature. 

\section{Hidden Ideal Equations}Instead of a single bivariate polynomial,
Alice may employ an ideal of a very modest size. She separates
the variables that she
employs within two sets, $\{X_i\}$, $\{Y_j\}$; one for encryption, one
for decryption. She may decide to leave one of the equations employed
of higher degree in the $\{Y_j\}$ after manipulations, so she gives raise to a
probabilistic encryption protocol.\label{ideal} Alice obtains
her PK with manipulations as in section \ref{key} on all
variables $\{X_i\}$, $\{Y_j\}$. Her parameters are: 
\begin{itemize}
\item $n=[\mathbb{K}:\mathbb{F}_q]$;
\item the number $s_1$, $s_2$ of variables $\{X_i\}$, $\{Y_j\}$, respectively;
\item the number $r$ of private equations.
\end{itemize}

So, the number of PK equations is $n\cdot r$, the number of the
variables $x_{ij}$ is $n\cdot s_1$, and that of the $y_{kl}$ is
$n\cdot s_2$.

Alice's number of variables, the $\{X_i\}$, is insignificant so far, so she is
supposed to be able to appeal to Gr\"obner techniques in order to solve her
system of equations within the field of coefficients for Bob's
$\{Y_j\}$. 

What is most important here and throughout, if 
Bob succeeds to encrypt, Alice does always succeed to decrypt. 

For ease of treatment, assume now that Alice does not apply affine
transformations to her variables. Bob fails encryption for a certain
cleartext $(X_1,\dots X_{s_1})$ iff Alice's private ideal has no solutions
in the $Y$ for such an $(X_1,\dots X_{s_1})$. Alice's private ideal is a
random one. If she takes $r\leq s_2$, the probability that it has no
solutions is $\approx 0$, and $\approx 1$ for $r> s_2$. So, it
suffices that Alice takes $r\leq s_2$. The rare critical cases that
may supervene are faced simply changing alphabet.

With slight changes, this reasoning holds in the case that Alice
applies affine transformations, too. 

The real problem is indeed that the solutions to Alice may be too many; and in
any case finitely many, as the base field is finite. The best remedy
to that is that Alice takes $r=s_1$. So, the ideal that she obtains
after substitution of Bob's ciphertext is zerodimensional (quite easy
to cause it happen), and the number of solutions is bounded 
above by the total degree of the system. So, she can contain the
number of solutions by taking the total degree in the $\{X_i\}$ modest. 

Alice can take all equations of
very low degree in the $X$, and then transform that basis of the ideal
they generate to another one of very high degrees in the $X$. So she
has a low Bezout number of the ideal, and higher degrees in the $X$,
and transformations as above can take place.
If she takes the first basis linear, the number of solutions of her
equations reduce to one: Bob's cleartext. She can substitute Bob's
ciphertext to any of bases of her private ideal, e.g., to a linear one.

As soon as $r>s_1$, the PK becomes overdefined.

Alice applies a permutation to the equations and a renumeration to the
variables before publishing her key, so Eve does not know how are they
related. She may apply 
affine transformations, or may not, or may apply to only some of the
$X_i$, $Y_j$; at her discretion.

If $s_1< s_2$, the size of the ciphertext is
bigger than that of cleartext, and nothing else wrong. By this case,
encryption is practically always probabilistic. Indeed, even when the
equations are linear with respect to the $y_{kl}$, since there are more
variables than equations, the solutions exist, and are not unique.

Actually, Alice can take a big $s_2$. She may choose to
manipulate some of the $Y_j$ within a subfield of $\mathbb{K}$, rather than
within $\mathbb{K}$. Doing so, she is allowed a big $s_2$, and a
contained size of the ciphertext. The number of the variables $y_{kl}$
now is no more $n\cdot s_2$.

One can employ this protocol for signcryption. The sizes of
ciphertexts throughout are roughly equal to those of the plaintext
ones. So, one can use all the protocols we describe throughout for 
multiple encryption as well. They seem suitable for private key
schemes, too.  

Now the size of the PK is
$\mathcal{O}(s_1(n)^{t+1})$, and the complexity of the
trapdoor problem is $\mathcal{O}(t^{n\cdot s_1})$.

Even though the size of PK throughout grows polynomially with
$n$, before $n$ becomes interesting, the PK is already
quite cumbersome. 
So, opting for the choices of this section we can employ
much smaller $n$, whence moderate a lot the size of the public
key. 

Actually, $n=20$ or so is quite good. We 
are allowed some more values of $t$, too. Alice takes $s_1$ as big as
she can handle, e.g., $s_1=5, \ 6, \ 7$, or more. 

For $ns_1$ fixed, the bigger $s_1$ is, the exponentially less cumbersome the
PK is, and the exponentially harder becomes Eve's  task.

Generally speaking, Alice's task becomes exponentially harder with $s_1$,
too. In practice, it depends very much on whether does she have any
good basis of her private ideal, or not. In any case, the
speeds of becoming harder of tasks of Alice and Eve are quite different.

\subsection{}There exist classes of ideals called {\it with doubly
  exponential ideal membership property} \cite{swanson}. These are the
  ideals for which the calculus of a Gr\"obner basis requires  doubly
  exponential time on the number of variables. It is very 
  interesting to know whether can we employ them in some fashion in
  this class of cryptosystems. In any fashion, this is the theoretical
  limit for employing solving of polynomial systems of equations in
  PK cryptography.

\section{Some Considerations}
The idea of PK was first proposed by Diffie and Hellman
\cite{pkc}. Since then, it has 
seen several vicissitudes \cite{odlyzko, mora, mora2}.  

A trapdoor function is a map from cleartext units to ciphertext
units that can be feasibly computed by anyone having the
PK, but whose inverse function cannot be
computed without its knowledge:\begin{itemize}
\item either because (at present, publicly)
  there is no known way; 
\item or there are, but the amount of calculi is
  deterring.\end{itemize}   

Shannon \cite{stinson} called {\it unconditionally secure
  cryptosystems} those with trapdoor of the first class.  

Actually, the aim is to render the trapdoor problems equivalent to
  time-honoured hard 
  mathematical problems. Being of a problem hard or
  undecidable implies 
  nothing  a priori about the security of a cryptosystem
  \cite{odlyzko}, however.

Recall that of all schemes ever set up, only two of
  them, $RSA$ \cite{rsa} and {\it ECDL} \cite{koblitz},
  are going to be broken (or, at least, are going to become
  impractical) by solving their hard problems. 

The author is very fond of the idea of the PK, and
believes howsoever in new developments that will make it fully suffice
for all purposes.

Actually, one tendency is that of investigating {\it poor
  structures}, mean, structures with less operations, like groups,
semigroups with cryptosystems upon the {\it word problem}
  \cite{anshel, yamamura, hughes}. Yamamura's paper \cite{yamamura}
  can be considered a pioneering USPK. Unfortunately, its scheme is
  still uneffective. 
 
William Sit and the author are investigating {\it rich
  structures}. We are investigating among other things effective 
  USPK schemes upon 
differential fields of positive characteristic. We
hope that cryptography will arouse new interests on differential and
universal algebra, too, as it did in number theory and arithmetic
geometry. One reason of optimism is that in universal algebra one can
go on further with new structures and hard or undecidable problems
forever. Until now we have appealed 
to only unary and binary arithmetic operations.
\section{Generalizations on Differential Fields}
Differential\footnote{Most of considerations given in this section are
  suggestions of professor Sit through private communications.}
  algebra \cite{kolchin, sit2, ritt, sadik, kaplansky} 
  owes its existence mostly to the efforts of Ritt  
\cite{ritt} to handle differential equations by means of
algebra. 

A differential field is a field $\mathbb{F}$ endowed with a
  set of linear maps $\theta : \mathbb{F}\longrightarrow\mathbb{F}$ called
  derivatives, such that: $\theta(ab)=a\theta(b)+\theta(a)b$. 

Kaplansky's booklet is perhaps the best introduction in the topic.

The schemes given throughout work as well in
differential settings. Take 
$\mathbb{K}$ to be a finite
differential field extension of a differential field
$\mathbb{F}$ of positive characteristic\footnote{In zero
  characteristic numerical analysis tools seriously affect security,
  or at least constrain us to more careful choices. We shall
  not dwell on this topic here.}.
Any such $\mathbb{K}$ is defined by a system of linear homogeneous
differential equations, and there are structural constants defining
the operations for the derivations (one matrix for each derivation),
as well for multiplication. 

One can now replace (\ref{poly1}) with a differential polynomial of
higher order and degree. Throughout section \ref{ideal}, one can
replace ideals with small suitable differential ideals, too. The
schemes work verbatim. 

The techniques given throughout for polynomials, if
applied to differential polynomials, will definitely make it much harder
to attack any protocol developed. Any affine transformation (by this is
meant a linear combination of the differential indeterminates with
not-necessarily constant coefficients, and this linear combination is
then substituted  {\bf differentially}  in place of the differential
indeterminates) will not only even out the degrees, but also the orders
of the various partials, and making the resulting differential
polynomials very dense. 

However, there is one thing to caution about:
any time one specifies these structural matrices, they have to satisfy
compatibility equations. In the algebraic case, it is the relations
between $P^k=\{{p_{ij}}^{(k)}\}$ in (\ref{id3}) and
$M_{\ell}=\{m_{ij\ell}\}$ in (\ref{id4}). The $P^k$ are simply determined
uniquely by $M_{\ell}$, given the choices implicitely defined in (\ref{id4}).

It is very interesting to know in the algebraic case whether Alice's
PK is invariant under a change of basis, all the other settings
being equal. There is probably some group of matrices in $GL(n, q)$
that can do that. Such a knowledge would only weaken all cryptosystems
based on equations systems solving. 

In the differential case there is a similar action called Loewy
action, or the gauge transformation. For ordinary differential
equations, two matrices $A$, $B$ are Loewy similar if there is an
invertible matrix $K$ such that $A=\delta K\cdot
K^{-1}+KBK^{-1}$. Using this action, one can classify the different
differential vector space structures of a finite dimensional vector
space. There is also a cyclic vector algorithm to find a special basis,
so that the differential linear system defining the vector space
becomes equivalent to a single linear $ODE$. 

If no other problems arise for the differential
algebraic schemes, there is however
one caution more for them to be unconditionally secure. We have to avoid the
exhaustive search. For that, Alice has to publish a finite alphabet
where each letter is represented by an infinite set, disjoint sets for
different letters. This is possible in differential fields, as
they are infinite. Alice renders the sets public parametrically, as
differential algebraic functions of elements of the base differential
field, and parameters, e.g., in $\mathbb{Z}$. Bob
chooses a letter, gives random values to parameters, obtains one
representant of the letter, and proceeds as above. In any case, if
$\mu$ is the order of public equations, any two elements $\Xi$, $\Theta
\in \mathbb{F}$ such that $(\Xi - \Theta)^{(\mu )}=0$ must represent
the same letter, if any. 

In the algebraic case such constructions do not make sense, as the base
field is finite. Besides, Gr\"obner attack is always at hand.

The main care for Alice is that the PK
equations must not fall into feasible cases by well-known means,
such as linear algebra. 

Now the size of the PK is $\mathcal{O}(n^{to+1})$,
where $o$ is the order of PK equations. Quite
explosive!!! One more reason to take $q=2$, so some more
monomials reduce to zero. 

Anyway, we do not have to
increase parameters for better security. The trapdoor problem is
simply undecidable. 
Unlike the algebraic case, we can split cleartext into small strings. 
Actually, quite good sample values are: $n=20$ and $t, o=2, 3, 4$, or so.
As of now, $HDPE$ trapdoor problem seems undecidable, and the
scheme effective. 
\subsection*{Acknowledgments.}
I wish to thank  Massimiliano Sala and Christopher Wolf
for many suggestions and fruitful discussions. I am
particularly indebted to William Sit for several comments and
improvements on earlier drafts, and to his advisor, Carlo Traverso.

\addcontentsline{toc}{section}{Bibliography}
\bibliographystyle{alpha}
\bibliography{biblio}
\nocite{HFE, Patarin95, gathen, odlyzko, koblitz,
  marcus, moh, imai1, imai2, sit,  patarin96hidden, pkc, sadik,
  kolchin, sit2, ritt, hughes, anshel, yamamura, gathen, stinson,
  ckps, patarin96hidden, Wolf:02:Thesis, menezes, swanson, mora, mora2}

\end{document}